\documentclass[letter]{IEEEtran}

\usepackage{kotex}
\usepackage{ulem}

\ifCLASSINFOpdf
\else
   \usepackage[dvips]{graphicx}
\fi
\usepackage{url}
\usepackage{cite}
\usepackage{lipsum} 
\usepackage{graphicx}
\usepackage{amsmath, amsfonts}
\usepackage{xcolor}
\usepackage{booktabs, multirow}
\ifCLASSOPTIONcompsoc
    \usepackage[caption=false, font=normalsize, labelfont=sf, textfont=sf]{subfig}
\else
\usepackage[caption=false, font=footnotesize]{subfig}

\newcommand{\ie}{\textit{i}.\textit{e}., }

\newcommand{\mb}[1]{\mathbf{#1}}

\begin{document}

\title{Split-KalmanNet: A Robust Model-Based Deep Learning Approach for SLAM}

\author{Geon Choi, \IEEEmembership{Graduate Student Member, IEEE}, Jeonghun Park,\IEEEmembership{ Member, IEEE}, Nir Shlezinger, \IEEEmembership{ Member, IEEE},\\Yonina C. Eldar \IEEEmembership{Fellow, IEEE} and Namyoon Lee, \IEEEmembership{Senior Member, IEEE}
\thanks{G. Choi is with the Department of Electrical Engineering, POSTECH, Pohang, Gyeongbuk, 37673 Korea
(e-mail: {simon03062}@postech.ac.kr).}
\thanks{J. Park is with the School of Electronics Engineering, College of IT Engineering, Kyungpook National University, Daegu 41566, South Korea
(e-mail: jeonghun.park@knu.ac.kr).}
\thanks{N. Shlezinger is with the School of Electrical and Computer Engineering, Ben-Gurion University of the Negev, Beer-Sheva, Israel (e-mail: nirshl@bgu.ac.il).}
\thanks{Y. C. Eldar is with the Math and CS Faculty, Weizmann Institute of Science, Rehovot, Israel (e-mail: yonina.eldar@weizmann.ac.il).}

\thanks{N. Lee is with the School of Electrical Engineering, Korea University, Seoul 02841, South Korea (e-mail: namyoon@korea.ac.kr).}}

\maketitle

\begin{abstract}
Simultaneous localization and mapping (SLAM) is a method that constructs a map of an unknown environment and localizes the position of a moving agent on the map simultaneously. Extended Kalman filter (EKF) has been widely adopted as a low complexity solution for online SLAM, which relies on a motion and measurement model of the moving agent. In practice, however, acquiring precise information about these models is very challenging, and the model mismatch effect causes severe performance loss in SLAM. In this paper, inspired by the recently proposed KalmanNet, we present a robust EKF algorithm using the power of deep learning for online SLAM, referred to as Split-KalmanNet. The key idea of Split-KalmanNet is to compute the Kalman gain using the Jacobian matrix of a measurement function and two recurrent neural networks (RNNs). The two RNNs independently learn the covariance matrices for a prior state estimate and the innovation from data. The proposed split structure in the computation of the Kalman gain allows to compensate for state and measurement model mismatch effects independently. Numerical simulation results verify that Split-KalmanNet outperforms the traditional EKF and the state-of-the-art KalmanNet algorithm in various model mismatch scenarios. 
\end{abstract}


\section{Introduction}

Simultaneous localization and mapping (SLAM) is a method that jointly estimates  the location of a moving agent and the map of an unknown environment \cite{chatila1985position}. This method was initially studied in the context of autonomous control of robots, and has been extended to various applications, including unmanned aerial vehicles (UAV), self-driving cars, and augmented reality (AR) \cite{caballero2009vision, taketomi2017visual, chen2020survey}. However, performing precise SLAM is very challenging in practice. The main difficulty arises from the fact that building a map of the unknown environment requires perfect knowledge of the position information of a moving agent. Furthermore, accurate localization also requires exact map information. To apply SLAM, three models are needed: i) a motion model of a moving agent, ii) an inverse measurement model that determines the locations of landmarks (or features) from measurement data, and iii) a measurement model that predicts the agent's location from estimated landmark positions.



Standard approaches for SLAM are based on the Kalman filter (KF), particle filter, and graph-based methods \cite{smith1986representation, cappe2007overview, grisetti2010tutorial}. In particular, the extended Kalman filter (EKF) is the most popular model-based algorithm for online SLAM \cite{smith1986representation}. The EKF method exploits various model knowledge obtained from the underlying physical dynamics in SLAM. Specifically, it alternately performs i) state update according to a predefined motion model of the agent and ii) measurement update according to a measurement model that acquires information for the relative ranges and angles from the agent to the landmarks in each time slot. The EKF is mathematically tractable and can achieve  optimal estimation under some mild conditions (e.g., linear and Gaussian process). This model-based online optimization method, however, is vulnerable to {the physical and statistical} model mismatch effects, and it may diverge when these effects are pronounced \cite{einicke1999robust}. 

Recently, a model-based DNN-aided Kalman filter algorithm called KalmanNet was proposed in \cite{revach2022kalmannet}. The idea of KalmanNet is to learn the Kalman gain matrix using model-less deep neural networks (DNNs) \cite{2022}. The learned Kalman gain matrix is then plugged into the model-based state and measurement updates of the KF algorithm. Thanks to the power of deep learning, the trained Kalman gain matrix is relatively robust to model mismatch effects compared to the traditional KF. When optimizing the Kalman gain with the DNN, however, KalmanNet does not completely exploit the model-based structure in computing the Kalman gain matrix. In particular, when state and measurement model mismatch effects coexist, the model mismatch effects can be coupled, leading to  performance loss in the state estimation.

In this paper, we present a new model-based deep learning algorithm for SLAM referred to as {\it Split-KalmanNet.} The main idea of Split-KalmanNet is to independently train the covariance matrices for a prior state estimate and for the innovation using two DNNs in parallel. Then, the Kalman gain matrix is obtained by combining the learned covariance matrices with the knowledge of the Jacobian matrix of the measurement function according to the model-based Kalman gain update rule. Our split learning structure makes the algorithm robust to both the state and measurement model mismatch effects because it mutually decouples these effects by independently optimizing the two DNNs. From simulations, we show that the proposed Split-KalmanNet provides a considerable gain in terms of  mean-squared error (MSE) and standard deviation compared to existing model-based EKF and KalmanNet under both the state and measurement model mismatch effects.

The paper is organized as follows. In Section~\ref{sec::system model}, we present a system model and problem formulation for SLAM. Then, we briefly review conventional model-based EKF and KalmanNet algorithms for SLAM in Section~\ref{sec::conventional approaches}. Section~\ref{sec::S-KalmanNet} introduces the proposed Split-KalmanNet algorithm. Section~\ref{sec::results} shows the effectiveness of the proposed method for SLAM via simulations.

\section{System Model and Problem Formulation}\label{sec::system model}
In this section, we explain a SLAM problem. The goal of SLAM is to jointly estimate the position of a moving agent and landmarks using relative range and angle information between the agent and landmarks. This SLAM system can be modeled as a discrete-time nonlinear dynamical system comprised of two models: i) a state evolution model and ii) a measurement model. We shall briefly explain these models according to \cite[Chapter~10]{thrun2002probabilistic}.

{\bf State evolution model:} We denote the position of a moving agent and the moving-direction angle at time $t$ by $\left(x_t^{\sf R}, y_t^{\sf R}\right)$ and $\theta_t^{\sf R}$. Further, we define the position of the $m$th landmark as $\left(x_m^{\sf L}, y_m^{\sf L}\right)$. By concatenating all these variables, the state vector at time $t$ is defined as
\begin{align}
\mb{x}_t &=\left[x_t^{\sf R}, y_t^{\sf R}, \theta_t^{\sf R}, x_1^{\sf L}, y_1^{\sf L},  \ldots,  x_M^{\sf L}, y_M^{\sf L}\right]^{\top}\in \mathbb{R}^{3+2M}. \label{eq:state}
\end{align}
The agent is assumed to move in an unknown environment according to a motion model with a constant velocity $v_t^{\sf R}$ and rotation angle $d\theta_t^{\sf R}$, namely, 
\begin{align}
    x_{t+1}^{\sf R} &= x_t^{\sf R} + v_t^{\sf R} \cos \theta_t^{\sf R} + w_{t}^{x}, \nonumber \\
    y_{t+1}^{\sf R} &= y_t^{\sf R} + v_t^{\sf R} \sin \theta_t^{\sf R} + w_{t}^{y}, \nonumber \\
    \theta_{t+1}^{\sf R} &= \theta_t^{\sf R} + d\theta_t^{\sf R} + w_{t}^{\theta}, \label{eq:model_move}
\end{align}
where $w_{t}^{x}$, $w_{t}^{y}$, and $w_{t}^{\theta}$ denote process noise in $x$-axis, $y$-axis, and the rotation angle. The distributions of $w_{t}^{x}$, $w_{t}^{y}$, and $w_{t}^{\theta}$ are assumed to be independent zero-mean Gaussian. Using \eqref{eq:model_move}, the state vector ${\bf x}_t$ evolves to ${\bf x}_{t+1}$ at time $t+1$ with a nonlinear mapping function ${\bf f}:\mathbb{R}^{3+2M}\times \mathbb{R}^2 \rightarrow \mathbb{R}^{3+2M}$ as
\begin{align}
    {\bf x}_{t+1} = {\bf f}\left( {\bf x}_t , v_t^{\sf R}, d\theta_t^{\sf R} \right)  +{\bf w}_t, \label{eq:vector motion model}
\end{align}
 with process noise vector $
    \mb{w}_t = 
    \begin{bmatrix}
    w_{t}^{x} & w_{t}^{y} & w_{t}^{\theta} & {\bf 0}_{2M}
    \end{bmatrix}^{\top}.$
The time-invariant  covariance matrix of ${\bf w}_t$ is defined as $ {\bf Q}=\mathbb{E}\left[ {\bf w}_t {\bf w}_t^{\top}\right]$.

{\bf Measurement model:}
For each time $t$, the moving agent is assumed to acquire the range and angle measurements from $M$ landmarks. Let $r_{m,t}$ and $\phi_{m,t}$ be the range and angle measurements from the $m$th landmark at time $t$. Then, these measurements are given by 
\begin{align}
    r_{m,t} &= \sqrt{\left(x_{m}^{{\sf L}} - x_t^{{\sf R}}\right)^2 + \left(y_{m}^{\sf L} - y_t^{\sf R}\right)^2}, \nonumber\\
    \phi_{m,t} &= \operatorname{atan2}\left(y_{m}^{\sf L} - y_t^{\sf R}, x_{m}^{\sf L} - x_t^{\sf R}\right) - \theta_t^{\sf R},\label{eq:measurement}
\end{align} for $m\in [M]$. Here, $\operatorname{atan2}\left(y, x\right)$ denotes the angle between the $x$-axis and the point $(x, y)$ in radians in the Euclidean plane. By concatenating all these measurements, we define a measurement vector at time $t$ as 
\begin{align}
{\bf h}({\bf x}_t) &= \left[r_{1,t}, \phi_{1,t}, \ldots, r_{M,t}, \phi_{M,t}\right]^{\top} \in \mathbb{R}^{2M},
\end{align}
where ${\bf h}(\cdot):\mathbb{R}^{3+2M}\rightarrow \mathbb{R}^{2M}$ denotes a nonlinear measurement function from state vector ${\bf x}_t$ to the $2M$ measurements in \eqref{eq:measurement}. As a result, the noisy measurement vector at time $t$ is given by
\begin{align}
    {\bf y}_t = {\bf h}({\bf x}_t) +{\bf v}_t, \label{eq:vector measurement model}
\end{align}
where ${\bf v}_t$ denotes the measurement noise vector at time $t$ whose distribution follows a zero-mean Gaussian.
Under the premise that the measurement noise is independent and identically distributed over time index and landmarks, the covariance matrix is defined as ${\bf R} = {\mathbb E} \left[{\bf v}_t {\bf v}_t^{\top}\right] = {\bf I}_{M\times M} \otimes {\bf D}_{2 \times 2}^{\bf R}$. Here, ${\bf D}_{2 \times 2}^{\bf R}$ is the noise covariance matrix of the range and angle measurements.

{\bf Online SLAM optimization:}
The online SLAM optimization problem is equivalent to the sequential minimum mean squared error (MMSE) estimation problem for each time $t\in [T]$ with known the state transition function ${\bf f}(\cdot)$ and  measurement function ${\bf h}(\cdot)$ :
\begin{equation}
    \operatorname*{arg\,min}_{\hat{\mb{x}}_t}\mathbb{E}\left[\left(\mb{x}_t - \hat{\mb{x}}_t \right)^2 \vert ~ \mb{y}_1, \ldots, \mb{y}_t \right]. \label{eq:optimization}
\end{equation}
The EKF algorithm provides a recursive solution for this problem. When implementing the EKF algorithm, it is required to know both process and measurement noise distributions accurately. For instance, under the zero-mean Gaussian noise assumption, the precise knowledge of the covariance matrices ${\bf Q}$ and ${\bf R}$ is needed. In practice, however, acquiring perfect knowledge of this statistical information is difficult.
{This model mismatch effect causes severe performance loss in the EKF algorithm. Consequently, a robust sequential state estimation algorithm is required for the model mismatch problem.}

\section {Conventional Approaches}\label{sec::conventional approaches}
In this section, we briefly review the EKF and KalmanNet algorithm in \cite{revach2022kalmannet} to solve the online SLAM problem in \eqref{eq:optimization}.

\subsection{Model-Based Extended Kalman Filter (MB-EKF)}
The EKF algorithm performs i) state-update and ii) measurement update recursively. To present the algorithm concisely, we first define some notations. Let {\it a priori} mean and covariance of the state vector at time $t$ be ${\bf \hat x}_{t\vert t-1} $ and ${\bf \Sigma}_{t\vert t-1} = \mathbb{E}\left[ ({\bf x}_t-{\bf \hat x}_{t\vert t-1})({\bf x}_t-{\bf \hat x}_{t\vert t-1})^{\top}\right]$. We also denote {\it a posteriori} mean and its covariance at time $t$ by ${\bf \hat x}_{t\vert t}$ and ${\bf \Sigma}_{t\vert t} = \mathbb{E}\left[ ({\bf x}_{t}-{\bf \hat x}_{t\vert t})({\bf x}_{t}-{\bf \hat x}_{t\vert t})^{\top}\right]$.

{\bf State-update:} The state-update step predicts {\it a priori} mean and covariance of the state vector using {\it a posteriori} mean ${\bf \hat x}_{t-1|t-1}$ and covariance matrix ${\bf \Sigma}_{t-1\vert t-1}$, which are obtained in the previous measurement-update step:
\begin{align}
    {\bf \hat x}_{t\vert t-1} &= {\bf f}\left( {\bf \hat x}_{t-1|t-1}, v_t^{\sf R}, d\theta_t^{\sf R}\right), \nonumber \\
    {\bf \Sigma}_{t\vert t-1} &= {\bf F}_{t-1}{\bf \Sigma}_{t-1\vert t-1}{\bf F}_{t-1}^{\top} +{\bf Q},\label{eq:sateupdate}
\end{align}
where ${\bf F}_{t}  =\frac{\partial {\bf f}}{\partial {\bf x}}|_{{\bf x}={\bf \hat x}_{t|t}}$ is the Jacobian of ${\bf f}(\cdot)$ evaluated at ${\bf \hat x}_{t|t}$.

{\bf Measurement-update:} In the measurement-update step, {\it a posteriori} mean ${\bf \hat x}_{t|t}$ and covariance matrix ${\bf \Sigma}_{t\vert t}$ are computed using  {\it a priori} mean ${\bf \hat x}_{t\vert t-1}$ and covariance matrix $ {\bf \Sigma}_{t\vert t-1}$ attained in the previous state-update step. In this filtering step, the mean of measurement ${\bf \hat y}_{t\vert t-1} = {\bf h}\left({\bf x}_{t\vert t-1}\right)$ is computed. Then, the Kalman gain matrix refines {\it a priori} mean ${\bf \hat x}_{t\vert t-1}$ using the innovation ${\bf y}_t - {\bf \hat y}_{t\vert t-1}$ acquired in time $t$ toward minimizing the MSE. The Kalman gain matrix is given by
\begin{align}
    {\bf K}_t = {\bf \Sigma}_{t\vert t-1}{\bf H}_t^{\top}{\bf S}_t^{-1}, \label{eq:kalmangain}
\end{align}
where ${\bf H}_t=\frac{\partial {\bf h}}{\partial {\bf x}}|_{{\bf x}={\bf \hat x}_{t|t-1}}$ is the Jacobian of ${\bf h}(\cdot)$ evaluated at ${\bf \hat x}_{t|t-1}$ and ${\bf S}_t=\mathbb{E}\left[({\bf y}_t - {\bf \hat y}_{t\vert t-1})({\bf y}_t - {\bf \hat y}_{t\vert t-1})^{\top}\right]$ is the covariance matrix of the innovation. Using the Kalman gain in \eqref{eq:kalmangain}, {\it a posteriori} mean and covariance are updated as
\begin{align}
    {\bf \hat x}_{t\vert t} &=  {\bf \hat x}_{t\vert t-1} + {\bf K}_t \left( {\bf y}_t - {\bf \hat y}_{t\vert t-1}\right), \nonumber \\
    {\bf \Sigma}_{t\vert t} &=   {\bf \Sigma}_{t\vert t-1}  - {\bf K}_{t}{\bf S}_{t}{\bf K}_{t}^{\top}. \label{eq:measurementupdate}
\end{align}
{The performance of this model-based EKF algorithm degrades severely in the presence of model mismatch effects.}

\subsection{KalmanNet - A Hybrid Approach}\label{sec::KalmanNet}
The KalmanNet \cite{revach2022kalmannet} is a hybrid approach that exploits the power of a DNN while maintaining the recursive EKF algorithm structures in \eqref{eq:sateupdate}, \eqref{eq:kalmangain}, and \eqref{eq:measurementupdate}. The noticeable difference from the model-based EKF algorithm is in computing the Kalman gain matrix. Instead of using the Kalman gain matrix in \eqref{eq:kalmangain} for the measurement-update, KalmanNet performs the measurement-update using a learned Kalman gain matrix, denoted by ${\mathcal K}_t(\Theta)$, with trainable parameters $\Theta\in \mathbb{R}^{N}$, \ie
\begin{align}
     {\bf \hat x}_{t\vert t} =  {\bf \hat x}_{t|t-1} + {\mathcal K}_t(\Theta) \left( {\bf y}_t - {\bf \hat y}_{t|t-1}\right).
\end{align}
The Kalman gain is trained by optimizing the squared-error loss function with respect to the parameters $\Theta$ in an end-to-end fashion. 

Let $\mathcal{D}=\left\{ ({\bf X}^{(\ell)}, {\bf Y}^{(\ell)})\right\}_{\ell=1}^L$ be a dataset comprised of $L$ different trajectories of the state-measurement pairs. The $\ell$th trajectory contains $T_{\ell}$ data samples $({\bf X}^{(\ell)}, {\bf Y}^{(\ell)})$ with ${\bf X}^{(\ell)}=\left[{\bf x}_1^{(\ell)}, {\bf x}_2^{(\ell)}, \ldots, {\bf x}_{T_{\ell}}^{(\ell)}\right]$ and ${\bf Y}^{(\ell)}=\left[{\bf y}_1^{(\ell)}, {\bf y}_2^{(\ell)}, \ldots, {\bf y}_{T_{\ell}}^{(\ell)}\right]$. We also define the empirical loss function for the $i$th training trajectory as
\begin{align}
    \mathcal{L}(\Theta) =\frac{1}{L} \sum_{\ell=1}^L\frac{1}{T_{\ell}} \sum_{t=1}^{T_{\ell}}\left\| {\bf x}_t^{(\ell)} - {\bf \hat 
    x}_{t|t}\left({\bf y}_t^{(\ell)}; \mathcal{K}_t(\Theta)\right)  \right\|_2^2. \label{eq:loss1}
\end{align} 
Let $\Delta {\bf x}_t^{(\ell)} =  {\bf x}_t^{(\ell)} -  {\bf \hat x}_{t|t-1}^{(\ell)}$ and $\Delta {\bf y}_t^{(\ell)} =  {\bf y}_t^{(\ell)} -  {\bf \hat y}_{t|t-1}^{(\ell)}$  be the state prediction error and the measurement innovation at time $t$ for trajectory $l$. The partial derivative of the loss function with respective to the Kalman gain matrix is
\begin{align}
   \frac{\partial  \mathcal{L}(\Theta)}{\partial \mathcal{K}_t(\Theta) }&=  \frac{1}{LT_{\ell}} \sum_{\ell=1}^L  \sum_{t=1}^{T_{\ell}}\frac{\partial \left\| \Delta {\bf x}_t^{(\ell)} -  \mathcal{K}_t(\Theta)\Delta{\bf y}_t^{(\ell)}  \right\|_2^2}{\partial \mathcal{K}_t(\Theta)}\nonumber\\
   &=\frac{2}{LT_{\ell}} \sum_{\ell=1}^L  \sum_{t=1}^{T_{\ell}} \left(\mathcal{K}_t(\Theta)\Delta {\bf y}_t^{(\ell)} -\Delta {\bf x}_t^{(\ell)}  \right)\left(\Delta {\bf y}_t^{(\ell)}\right)^{\top}. \label{eq:partial}
\end{align}
By plugging \eqref{eq:partial} into the following chain rule:
\begin{align}
    \frac{\partial  \mathcal{L}(\Theta)}{\partial \Theta} = \frac{\partial  \mathcal{L}(\Theta)}{\partial \mathcal{K}_t(\Theta) } \frac{\partial  \mathcal{K}_t(\Theta)}{\partial \Theta},
\end{align}
one can find a local optimal parameter $\Theta^{\star}$ such that $\frac{\partial  \mathcal{L}(\Theta^{\star})}{\partial \Theta^{\star}} = 0 $ using a stochastic gradient descent algorithm. To track the covariance matrices implicitly, recurrent neural networks (RNNs) were used to train the parameter using the innovation difference $\Delta {\bf y}_t= {\bf y}_t - {\bf \hat y}_{t|t-1}$ and the state update difference $\Delta {\bf \hat x}_t= {\bf \hat x}_{t|t} - {\bf \hat x}_{t|t-1}$ as input features. We refer to the detailed structure of the KalmanNet in \cite{revach2022kalmannet}.

This DNN-aided EKF algorithm does not require knowledge of the covariance matrices for process and measurement noise, \ie ${\bf Q}$ and ${\bf R}$. Rather, these matrices are implicitly learned from data for the Kalman gain computation; thereby, it can diminish statistical model mismatch effects compared to the model-based EKF method. In addition, it simplifies the state and time update procedures because {\it a priori} and {\it a posteriori} covariance matrices ${\bf \Sigma}_{t|t-1}$ and ${\bf \Sigma}_{t|t}$ can be discarded. Notwithstanding, this joint learning approach of the Kalman gain can lead to performance loss when one of  ${\bf Q}$ and ${\bf R}$ mismatch effects dominates the other. For instance, the dominant mismatch error in measurement ${\bf R}$ hinders learning the covariance matrix of states ${\bf Q}$. This fact motivates us to consider a split learning structure when computing the Kalman gain matrix.  



\section{Split-KalmanNet}\label{sec::S-KalmanNet}
In this section, we put forth a DNN-aided EKF algorithm, called Split-KalmanNet. To illustrate the proposed idea, it is instructive to rethink the Kalman gain matrix in the model-based EKF algorithm. As in \eqref{eq:kalmangain}, the Kalman gain matrix is computed by multiplying three factors: i) {\it a priori} covariance matrix of the state ${\bf \Sigma}_{t\vert t-1}$, ii) the Jacobian matrix of the measurement function ${\bf H}_t$, iii) the inverse of the covariance matrix of the innovation ${\bf S}_t^{-1}$. The first term, ${\bf \Sigma}_{t\vert t-1}$, is a function of both ${\bf Q}$ and ${\bf F}_t$. In addition, the last term, ${\bf S}_t^{-1}$, changes according to the second-order moment of the measurement noise ${\bf R}$. When the uncertainties in ${\bf Q}$ and ${\bf R}$ are heterogeneous, the end-to-end KalmanNet can be less robust because $\mathcal{K}_t(\Theta)$ cannot separate the statistical model mismatch effects in ${\bf Q}$ and ${\bf R}$.

\begin{figure}[t]
\centering
\includegraphics[width=1\columnwidth]{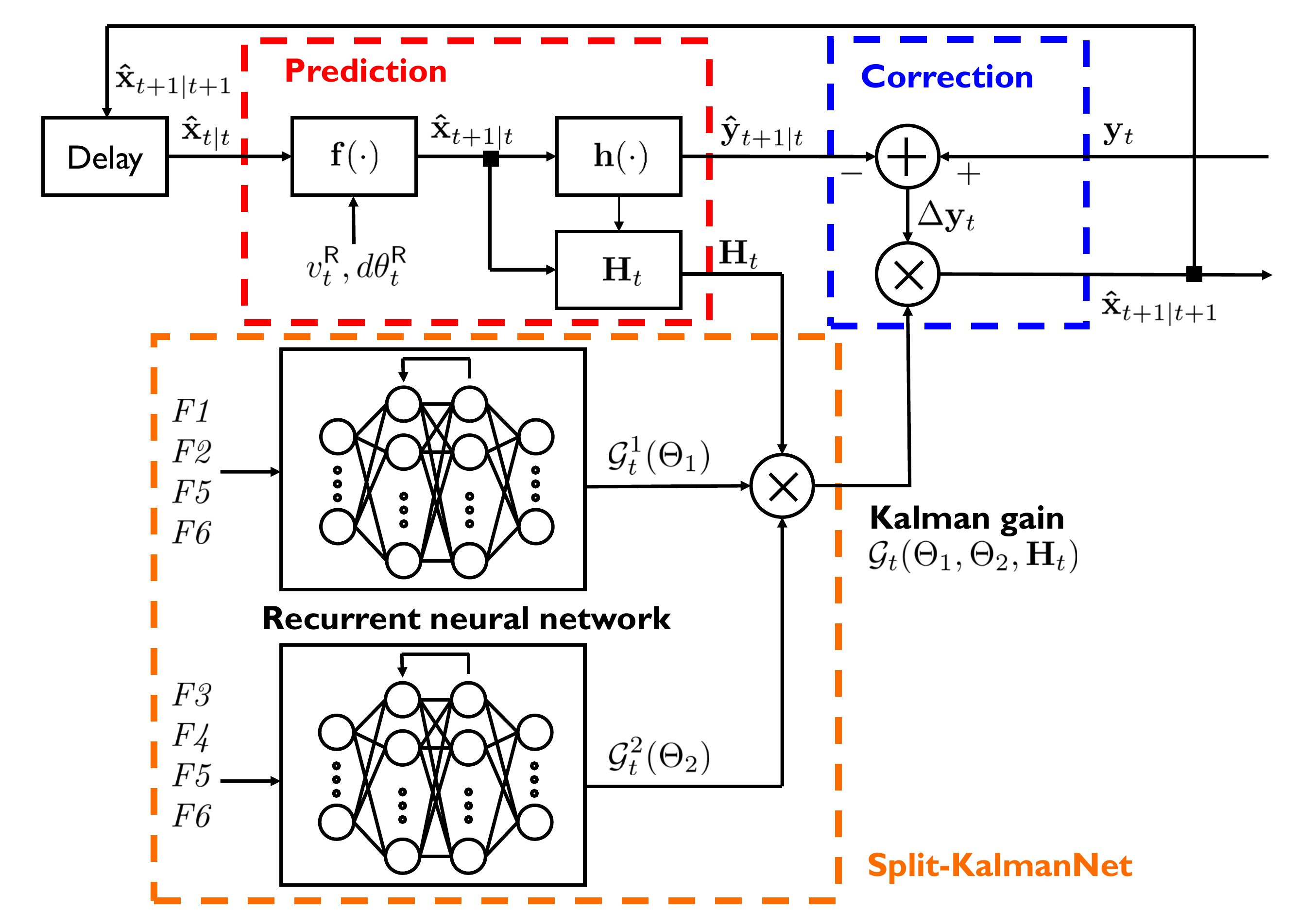}
\caption{Split-KalmanNet block diagram.}
\label{fig:split kalmannet}
\end{figure}

Harnessing the statistical independence property between the process and measurement noise vectors and the model-based knowledge in computing the Kalman gain matrix, our approach is to train the Kalman gain matrix with two separate DNNs to form the Kalman gain matrix:  
 \begin{align}
    \mathcal{G}_t(\Theta_1,\Theta_2, {\bf H}_t) = \mathcal{G}_t^1(\Theta_1){\bf H}_t\mathcal{G}_t^2(\Theta_2), \label{eq:kalmangain2}
\end{align}
where $\mathcal{G}_t^1(\Theta_1)$ is the DNN that learns {\it a priori} covariance matrix of the state ${\bf \Sigma}_{t\vert t-1}$ implicitly with parameter $\Theta_1\in \mathbb{R}^{N_1}$. In addition, $\mathcal{G}_t^2(\Theta_2)$ is another DNN that learns the inverse of the covariance matrix of the innovation ${\bf S}_t$ implicitly with parameter $\Theta_2\in \mathbb{R}^{N_2} $. Using this trained Kalman gain matrix, the Split-KalmanNet performs the state-update as
\begin{align}
     {\bf \hat x}_{t\vert t} =  {\bf \hat x}_{t|t-1} +   \mathcal{G}_t(\Theta_1,\Theta_2, {\bf H}_t) \left( {\bf y}_t - {\bf \hat y}_{t|t-1}\right).
\end{align}
The overall structure of Split-KalmanNet is illustrated in Fig.~\ref{fig:split kalmannet}.

The Split-KalmanNet allows independently learning the model uncertainties caused by the state update $({\bf F}_t, {\bf Q})$ and the measurement update $({\bf H}_t, {\bf R})$. Furthermore, the Split-KalmanNet decouples the effect of aforementioned model uncertainties from the Jacobian matrix of the measurement function ${\bf H}_t$ in computing \eqref{eq:kalmangain2}. As a result, it is more robust than the end-to-end KalmanNet in the presence of  heterogeneous model mismatch effects thanks to the split learning structure.

{\bf Loss function:} In analogy to \eqref{eq:loss1}, given dataset $\mathcal{D}=\left\{ ({\bf X}^{(\ell)}, {\bf Y}^{(\ell)})\right\}_{\ell=1}^L$,  we define a loss function in terms of two parameter sets $\Theta_1$ and $\Theta_2$:
\begin{align}
    \mathcal{L}_2(\Theta_1,\Theta_2) \!=\!\frac{1}{L}\! \sum_{\ell=1}^L\!\frac{1}{T_{\ell}} \!\sum_{t=1}^{T_{\ell}}\!\left\| {\bf x}_t^{(\ell)}\!\!\!-\!{\bf \hat 
    x}_{t|t}\!\left(\!{\bf y}_t^{(\ell)} \!; \mathcal{G}_t(\Theta_1,\Theta_2, {\bf H}_t)\!\right)\!\right\|_2^2. \label{eq:loss2}
\end{align} 

{\bf Alternating optimization for training:} We also provide an alternating optimization method to train two DNNs. We start by computing the partial derivative of the loss function with respective to the Kalman gain matrix:
\begin{align}
  & \frac{\partial  \mathcal{L}_2(\Theta_1,\Theta_2)}{\partial \mathcal{G}_t(\Theta_1,\Theta_2, {\bf H}_t) } \nonumber\\
   &=  \frac{1}{LT_{\ell}} \sum_{\ell=1}^L\sum_{t=1}^{T_{\ell}}\frac{\partial \left\| \Delta {\bf x}_t^{(\ell)} -  \mathcal{G}_t(\Theta_1,\Theta_2, {\bf H}_t)\Delta{\bf y}_t^{(\ell)}  \right\|_2^2}{\partial \mathcal{G}_t(\Theta_1,\Theta_2, {\bf H}_t)}\nonumber\\
   &=\frac{2}{LT_{\ell}} \sum_{\ell=1}^L\sum_{t=1}^{T_{\ell}} \left(\mathcal{G}_t(\Theta_1,\Theta_2, {\bf H}_t)\Delta {\bf y}_t^{(\ell)} -\Delta {\bf x}_t^{(\ell)}  \right)\left(\Delta {\bf y}_t^{(\ell)}\right)^{\top}. \label{eq:partial2}
\end{align}
Then, for given ${\bar \Theta_2}$, we first optimize $\Theta_1$ by solving the first-order optimally condition:
\begin{align}
    \frac{\partial  \mathcal{L}_2(\Theta_1,{\bar \Theta_2})}{\partial \Theta_1}  
    = \frac{\partial  \mathcal{L}_2(\Theta_1,{\bar \Theta_2})}{\partial \mathcal{G}_t(\Theta_1,{\bar \Theta_2}, {\bf H}_t)}\frac{\partial     \mathcal{G}_t^1(\Theta_1)}{\partial \Theta_1}{\bf H}_t\mathcal{G}_t^2({\bar \Theta}_2)=0. \label{eq:optimize1}
\end{align}
Using the updated ${\bar \Theta_1}$, we optimize $\Theta_2$ such that 
\begin{align}
    \frac{\partial  \mathcal{L}_2({\bar \Theta}_1,{ \Theta_2})}{\partial \Theta_2}  = \frac{\partial \mathcal{L}_2({\bar \Theta}_1, {\Theta_2})}{\partial \mathcal{G}_t ({\bar \Theta_1}, {\Theta_2}, {\bf H}_t)} \mathcal{G}_t^1({\bar\Theta}_1){\bf H}_t \frac{\partial \mathcal{G}_t^2(\Theta_2)}{\partial \Theta_2}=0. \label{eq:optimize2}
\end{align}
Using the updated ${\bar \Theta}_2$, we again optimize $\Theta_1$ by solving \eqref{eq:optimize1}. We repeat \eqref{eq:optimize1} and \eqref{eq:optimize2} until the parameters converge.

{\bf Complexity:} Unlike KalmanNet, our algorithm requires ${\bf H}_t$ in each time. Fortunately, in the SLAM problem, the structure of ${\bf H}_t$ is fixed and it needs to be re-evaluated over time, which increases the complexity very marginally. 

{\bf Input features:} The DNN $\mathcal{G}_t^1(\Theta_1)$ learns {\it a priori} covariance matrix of the state ${\bf \Sigma}_{t\vert t-1}$ and the DNN $\mathcal{G}_t^2(\Theta_2)$ learns the inverse of the covariance matrix of the innovation ${\bf S}_t$ implicitly. To track these covariance matrices, we consider the following input features:
\begin{itemize}
    \item[\it F1] The {\it state update difference} $\Delta {\bf \hat x}_t= {\bf \hat x}_{t|t} - {\bf \hat x}_{t|t-1}$. The available feature at time $t$ is $\Delta {\bf \hat x}_{t-1}$.
    \item[\it F2] The {\it state evolution difference} $\Delta {\bf \tilde x}_t = {\bf \hat x}_{t\vert t} - {\bf \hat x}_{t-1 \vert t-1}$. The available feature at time $t$ is $\Delta {\bf \tilde x}_{t-1}$.
    \item[\it F3] The {\it innovation difference} $\Delta {\bf y}_t = {\bf y}_t - {\bf \hat y}_{t|t-1}$.     
    \item[\it F4] The {\it measurement difference} $\Delta {\bf \tilde y}_{t} = {\bf y}_{t} - {\bf y}_{t-1}$.
    \item[\it F5] The {\it linearization error} ${\bf h}\left({\bf \hat x}_{t\vert t-1}\right) - {\bf H}_t {\bf \hat x}_{t \vert t-1}$. 
    \item[\it F6] The {\it Jacobian matrix} of measurement function ${\bf H}_t$.
\end{itemize}
Features {\it F1} and {\it F3} characterize the uncertainty of state estimates, while features {\it F2} and {\it F4} characterize the information about the state and measurement evolution process. In addition, features {\it F5} and {\it F6} encapsulate the local behavior of the measurement function.





\section{Simulation Results}\label{sec::results}
This section presents simulation results to show the robustness of splitting architecture to the noise heterogeneity using uniform circular motion model and then presents simulation setting and results for online SLAM.

{\bf State estimation algorithm:}
We compare the following algorithms:
\begin{itemize}
    \item {EKF (perfect):} The model-based EKF for reference purpose. The exact model parameters are used.
    \item {EKF (mismatch):} The EKF using predetermined model parameters.  
    \item {KalmanNet:} The KalmanNet using input features \{{\it F1}, {\it F2}, {\it F3}, {\it F4}\} as presented in \cite{revach2022kalmannet}.
    \item {Split-KalmanNet:} The proposed Split-KalmanNet. The input features \{{\it F1}, {\it F2}, {\it F3}, {\it F4}, {\it F5}, {\it F6}\} are used.
\end{itemize}

{\bf Test metric:}
We use the mean-squared error (MSE) as the test metric, which is defined as
\begin{equation}
     \mathrm{MSE ~[dB]} = 10\log_{10}\left( \frac{1}{LT_{\ell}}\sum_{\ell=1}^L\sum_{t=1}^{T_{\ell}} \Vert \mb{x}_t - \hat{\mb{x}}_t \Vert^2 \right).
\end{equation}

\subsection{Uniform Circular Motion}
To clearly show the effect of splitting architecture to the noise heterogeneity, we consider the following simple two-dimensional state transition model:
\begin{align}
    {\bf x}_{t+1} = \begin{bmatrix}
    \cos\theta & -\sin\theta \\ \sin\theta & \cos\theta
    \end{bmatrix}{\bf x}_t + {\bf w}_t,
\end{align}
where ${\bf x}_t = \begin{bmatrix}x_t & y_t \end{bmatrix}^\top \in \mathbb{R}^2$ is the position vector in the Euclidean plane and $\theta$ is constant. For the measurement model, we consider i) linear model ${\bf h}({\bf x}_t) = {\bf x}_t$ and ii) nonlinear polar coordinate model ${\bf h}({\bf x}_t) = \begin{bmatrix}
    \Vert {\bf x}_t \Vert_2^2 & {\rm atan2}({\bf x}_t)
    \end{bmatrix}^\top$
where ${\rm atan2}({\bf x}_t)$ is the angle between the $x$-axis and the state vector ${\bf x}_t$ in radians in the Euclidean plane.
The noise covariance matrices are ${\bf Q}= \sigma_w^2{\bf I}$ and ${\bf R}= \sigma_v^2 {\bf I}$. To simulate the noise heterogeneity, we fix $\sigma_w^2=10^{-3}$ and control $\nu=\sigma_v^2 / \sigma_w^2$. 

\begin{figure}[t]
\centering
\includegraphics[width=1\columnwidth]{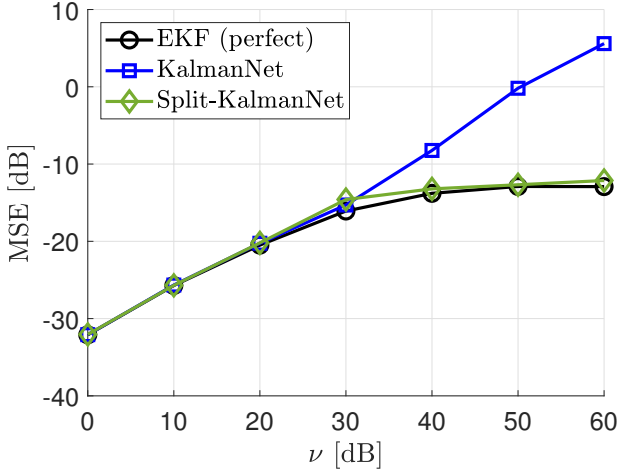} 
\caption{MSE (y-axis) comparison according to noise parameters (linear measurement).}
\label{fig:linear}
\end{figure}

The simulation result using linear measurement model is presented in Fig.~\ref{fig:linear}, which shows MSE of perfect EKF, KalmanNet, and our proposed Split-KalmanNet for various noise heterogeneity (i.e., $\nu$). When the noise heterogeneity is small, both the KalmanNet and Split-KalmanNet achieve the MSE of perfect EKF, and therefore MMSE. As the noise heterogeneity increases, however, the KalmanNet fails to learn the Kalman gain and shows higher MSE than that of perfect EKF, while our Split-KalmanNet achieves MMSE. It shows that our Split-KalmanNet is more robust to noise heterogeneity than other algorithms. 

\begin{figure}[t]
\centering
\includegraphics[width=1\columnwidth]{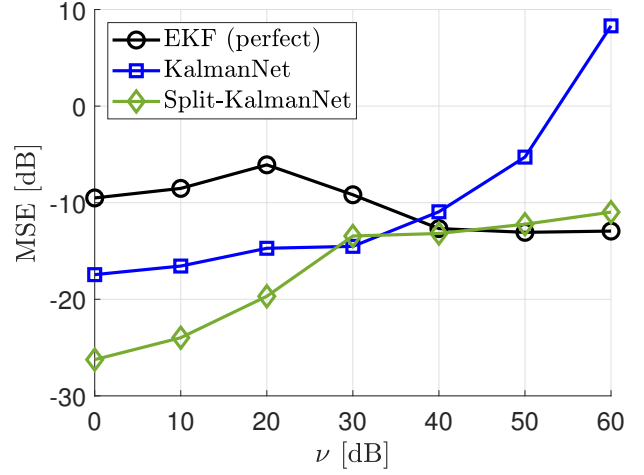} 
\caption{MSE (y-axis) comparison according to noise parameters (nonlinear measurement).}
\label{fig:nonlinear}
\end{figure}

Fig.~\ref{fig:nonlinear} presents the MSE of aforementioned algorithms in nonlinear measurement model for various $\nu$. For nonlinear measurement model, the perfect EKF does not achieve MMSE anymore and both the KalmanNet and Split-KalmanNet outperforms the perfect EKF when the noise heterogeneity is small. In particular, the Split-KalmanNet shows lower MSE than that of KalmamNet because the Split-KalmanNet further uses the Jacobian matrix of measurement function and linearization error to optimize the Kalman gain, while the KalmanNet does not. As the noise heterogeneity increases, the KalmanNet fails to learn the Kalman gain, while the Split-KalmanNet learns the Kalman gain and shows almost the same MSE compared to perfect EKF, which again confirms that our Split-KalmanNet is more robust to the noise heterogeneity than other algorithms.

\subsection{Online SLAM}\label{sec::simulation setting}
We compare the state estimation algorithms for online SLAM optimization. 

\begin{figure}[t]
\centering
\includegraphics[width=1\columnwidth]{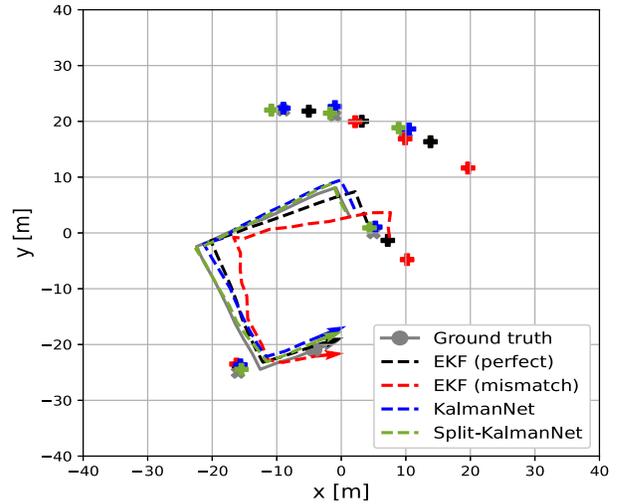} 
\caption{A realization of the estimated landmark positions and trajectories according to algorithms. Data is generated according to {\it D1}.}
\label{fig:env} \vspace{-0.2cm}
\end{figure}

{\bf Dataset generation:}
Each data in the dataset of size $L$ consists of the state and measurement vectors $({\bf X}^{(\ell)}, {\bf Y}^{(\ell)})$ of length $T_{\ell}$.
The environment consists of $M$ landmarks, where each landmark is uniformly placed on the integer-grid. 
Given the environment, the state and measurement vector are generated following the state evolution model in \eqref{eq:vector motion model} and the measurement model in \eqref{eq:vector measurement model}. In the simulation, we use two datasets configuration \{{\it D1}, {\it D2}\}, each of which is used for training and testing, respectively. The detailed parameters for the datasets are summarized in Table~\ref{tab:param}. Here, $q^2$ and $r^2$ are used to control the noise heterogeneity. For configuration {\it D2}, the parameters $\sigma_w^2$, $\sigma_v^2$, $q^2$, and $r^2$ will be specified later. Fig.~\ref{fig:env} visualizes estimated landmark positions and trajectories according to algorithms using dataset {\it D1}.

\begin{table}[t] 
\caption{Dataset summary}
    \vspace{-0.2cm}
    \centering
    \begin{tabular}{llll}
        \toprule
        {} & \multicolumn{2}{c}{Dataset} \\
        \cmidrule(lr){2-3} 
        {Parameters} & \multicolumn{1}{c}{$\mathcal{D}_1$ (Training)} & \multicolumn{1}{c}{$\mathcal{D}_2$ (Test)}  \\
        \midrule
        $(x_i^{\sf L}, y_i^{\sf L})$ [m] & $[-30,30]^2 \subset \mathbb{Z}^2$ &  $[-30,30]^2 \subset \mathbb{Z}^2$  \\
        $M$ & $5$ & $5$ \\
        $\sigma_w^2$ & $[5\times 10^{-4}, ~ 5 \times 10^{-2}]$ & -  \\
        $\sigma_v^2$ & $[5\times 10^{-4}, ~ 5 \times 10^{-2}]$ & -  \\
        ${\bf Q} / \sigma_w^2$ & ${\rm diag}(10, 10, 1, {\bf 0}_{2M})$ & ${\rm diag}(q^2, q^2, 1, {\bf 0}_{2M})$ \\
        ${\bf D}_{2\times 2}^{\bf R} / \sigma_v^2$ & ${\rm diag}(10^3, 1)$ & ${\rm diag}(r^2, 1)$  \\
        $v_t^{\sf R}$ [m/s] & $5$ & $1$  \\
        $d\theta_t^{\sf R}$ [radian] & $\mathcal{U}[-\pi, \pi]$ & $\mathcal{U}[-\pi, \pi]$  \\
        $L$ & $10^4$ & $10^3$  \\
        $T_{\ell}$ & $20$ & $50$  \\
        \bottomrule
    \end{tabular}
    \label{tab:param}
\end{table}

{\bf Test results:} 
{Using the trained model, we compare the MSE of each algorithm under various statistical model mismatch scenarios using dataset {\it D2}. For the purpose of reference algorithm under model mismatches, we adopt the EKF with predetermined noise parameters $q^2=10$, $r^2=10^2$, and $\sigma_v^2 = \sigma_w^2 = 10^{-3}$.} 

\begin{figure}[t]
\centering
\includegraphics[width=1\columnwidth]{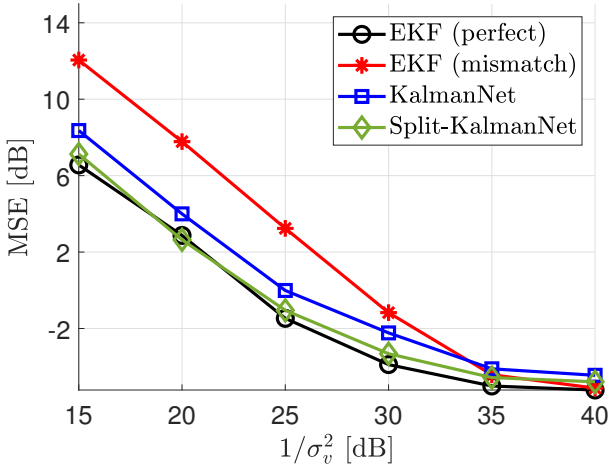}
\caption{MSE (y-axis) comparison according to the measurement noise variance.}
\label{fig:result 1} 
\end{figure}

First, we consider the effect of measurement noise variance $\sigma_v^2$. we set noise parameters to $q^2=10$, $r^2 = 10^3$, and $\sigma_w^2 = 10^{-3}$. For various $\sigma_v^2$, we present the MSE in Fig.~\ref{fig:result 1}. 
The proposed Split-KalmanNet achieves almost the same performance with perfect EKF for various noise parameters, while EKF using predetermined noise parameters shows performance degradation due to the statistical model mismatch.

\begin{figure}[t]
\centering
\includegraphics[width=1\columnwidth]{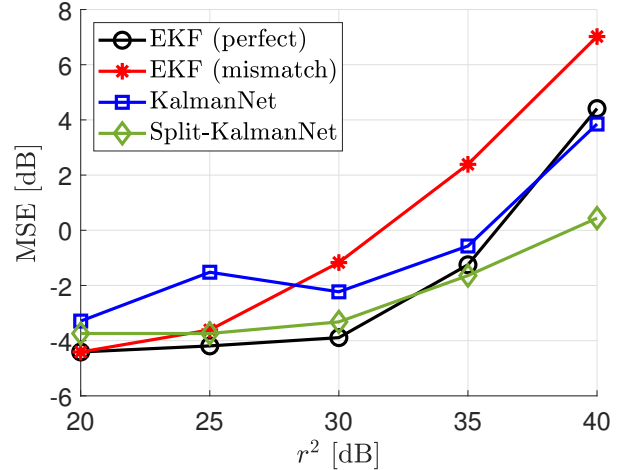} 
\caption{MSE (y-axis) comparison according to $r^2$.}
\label{fig:result 2}
\end{figure}

Second, we consider the noise heterogeneity. We set parameters to $\sigma_v^2 = \sigma_w^2 = 10^{-3}$, and $q^2=10$. For various $r^2$, we present the MSE in Fig.~\ref{fig:result 2}. 
Again, the proposed Split-KalmanNet achieves the almost same performance with perfect EKF.
{ In addition, it is noticeable that our proposed algorithm outperforms the perfect EKF by further exploiting the linear approximation error of the measurement function.}

\section{Conclusion}\label{sec::conclusion}
In this work, we presented a robust sequential state estimation algorithm for online SLAM optimization called {\it Split-KalmanNet}. The main idea of Split-KalmanNet was to train the covariance matrices for a prior state estimate and for the innovation using two DNNs in parallel. Leveraging the structure of the Kalman gain, Split-KalmanNet constructs the Kalman gain matrix using the trained covariance matrices with the Jacobian matrix of a measurement function. From simulations, we show that the proposed Split-KalmanNet provides a considerable gain in terms of MSE compared to existing methods under various model mismatch scenarios.

\normalem
\bibliographystyle{IEEEtran}
\bibliography{archive_v1}

\begin{thebibliography}{10}
\providecommand{\url}[1]{#1}
\csname url@samestyle\endcsname
\providecommand{\newblock}{\relax}
\providecommand{\bibinfo}[2]{#2}
\providecommand{\BIBentrySTDinterwordspacing}{\spaceskip=0pt\relax}
\providecommand{\BIBentryALTinterwordstretchfactor}{4}
\providecommand{\BIBentryALTinterwordspacing}{\spaceskip=\fontdimen2\font plus
\BIBentryALTinterwordstretchfactor\fontdimen3\font minus
  \fontdimen4\font\relax}
\providecommand{\BIBforeignlanguage}[2]{{%
\expandafter\ifx\csname l@#1\endcsname\relax
\typeout{** WARNING: IEEEtran.bst: No hyphenation pattern has been}%
\typeout{** loaded for the language `#1'. Using the pattern for}%
\typeout{** the default language instead.}%
\else
\language=\csname l@#1\endcsname
\fi
#2}}
\providecommand{\BIBdecl}{\relax}
\BIBdecl

\bibitem{chatila1985position}
R.~Chatila and J.-P. Laumond, ``Position referencing and consistent world
  modeling for mobile robots,'' in \emph{Proceedings. 1985 IEEE International
  Conference on Robotics and Automation}, vol.~2.\hskip 1em plus 0.5em minus
  0.4em\relax IEEE, 1985, pp. 138--145.

\bibitem{caballero2009vision}
F.~Caballero, L.~Merino, J.~Ferruz, and A.~Ollero, ``Vision-based odometry and
  {SLAM} for medium and high altitude flying {UAVs},'' \emph{Journal of
  Intelligent and Robotic Systems}, vol.~54, no.~1, pp. 137--161, 2009.

\bibitem{taketomi2017visual}
T.~Taketomi, H.~Uchiyama, and S.~Ikeda, ``Visual {SLAM} algorithms: A survey
  from 2010 to 2016,'' \emph{IPSJ Transactions on Computer Vision and
  Applications}, vol.~9, no.~1, pp. 1--11, 2017.

\bibitem{chen2020survey}
C.~Chen, B.~Wang, C.~X. Lu, N.~Trigoni, and A.~Markham, ``A survey on deep
  learning for localization and mapping: Towards the age of spatial machine
  intelligence,'' \emph{arXiv preprint arXiv:2006.12567}, 2020.

\bibitem{smith1986representation}
R.~C. Smith and P.~Cheeseman, ``On the representation and estimation of spatial
  uncertainty,'' \emph{The international journal of Robotics Research}, vol.~5,
  no.~4, pp. 56--68, 1986.

\bibitem{cappe2007overview}
O.~Capp{\'e}, S.~J. Godsill, and E.~Moulines, ``An overview of existing methods
  and recent advances in sequential {Monte Carlo},'' \emph{Proceedings of the
  IEEE}, vol.~95, no.~5, pp. 899--924, 2007.

\bibitem{grisetti2010tutorial}
G.~Grisetti, R.~K{\"u}mmerle, C.~Stachniss, and W.~Burgard, ``A tutorial on
  graph-based {SLAM},'' \emph{IEEE Intelligent Transportation Systems
  Magazine}, vol.~2, no.~4, pp. 31--43, 2010.

\bibitem{einicke1999robust}
G.~A. Einicke and L.~B. White, ``Robust extended {Kalman} filtering,''
  \emph{IEEE Transactions on Signal Processing}, vol.~47, no.~9, pp.
  2596--2599, 1999.

\bibitem{revach2022kalmannet}
G.~Revach, N.~Shlezinger, X.~Ni, A.~L. Escoriza, R.~J. Van~Sloun, and Y.~C.
  Eldar, ``Kalmannet: Neural network aided {Kalman} filtering for partially
  known dynamics,'' \emph{IEEE Transactions on Signal Processing}, vol.~70, pp.
  1532--1547, 2022.

\bibitem{2022}
N.~Shlezinger, Y.~C. Eldar, and S.~P. Boyd, ``Model-based deep learning: On the
  intersection of deep learning and optimization,'' \emph{arXiv preprint arXiv
  (https://arxiv.org/abs/2205.02640)}, 2022.

\bibitem{thrun2002probabilistic}
S.~Thrun, ``Probabilistic robotics,'' \emph{Communications of the ACM},
  vol.~45, no.~3, pp. 52--57, 2002.

\end{thebibliography}

\end{document}